# Effect of Landau-Zener tunneling by the varying sweeping rate of external field

J. B. Wang and G. P. Tong*

*College of Mathematics, Physics and Information Engineering, Zhejiang Normal University, Zhejiang Jinhua 321004, China*

We study the effect of Landau-Zener (LZ) tunneling caused by the varying sweeping rate of external field, formulating and approximately solving the problem with many levels of the LZ tunneling rate. Comparing with the steadily vary about sweeping field, the LZ tunneling rate will be essentially changed because of the unsteady variation of the sweeping field in time. Thus could help us to make the particles with in lower states transit periodically to upper states within the finite time.

As a basic quantum mechanical process, Landau-Zener (LZ) tunneling which concerns the particles in a crossing-level, will transit each other within a specific probability described by Landau and Zener [1,2]. Based on LZ theory, mass of literature has been devoted to its wide applications in real physical systems such as atoms in accelerating optical lattices [3], slow atomic and molecular collisions[4], field-driven superlattices [5] and molecular predissociation [6]. There has been much excellent work done by S. Brundobler *et al* [7-10] to solve the many levels problem although it would still be exactly solved. The generalization of this problem such as the variation of single particle level energies in multiple could affect the many particle state [11], that help us to understand the cavity QED experiment [12-16]. J. Keeling and V. Gurarie's work[17] on a photon mode in a cavity is also a helpful way to understand the Feshbach resonance, variation of the molecular energy caused by Feshbach resonance in cold atomic gases[11,18]. A. Altland and V. Gurarie generalized the many body of the LZ problem and found that adiabatic driving did not keep the system in its many particle ground state; rather, a finite fraction of particles remains in energetically high-lying states[11]. In addition, with the many body has been studied, the LZ tunneling for a nonlinear two-level energies which depend on the occupation of the levels within a periodic potential was done by Niu *et al* [19]. It helps us to understand the behavior of the LZ tunneling between Bloch bands of a Bose-Einstein condensate in an accelerated optical lattice deeply[20].

The aim of the present paper we proposed is to not only derive the transition probability which is different from the classical LZ model with the forms of sweeping field, but also to provide a theoretical understanding of the effect of the LZ tunneling caused by varying the sweeping rate of external field will transit with time-dependent. This circumstance corresponds to the one of the hot topics in modern nanophysics that the time-dependent external magnetic field induces the transition in the system in interacting spins [21,22]. As we know, in the linear case, the sweeping field $\gamma(t) = \alpha t$, by the way of solving for the eigenvalue and eigenstates of the Hamiltonian, we can derive the LZ transition in the near adiabatic regime $\alpha \to 0$. What we want to research is the case that the sweeping rate changes with time dependent, in particular, with periodically change that makes the system stay in adiabatic state periodically in some specific times. Although, the forms of the LZ tunneling probability in these two cases are similar, but the essential difference makes the physical property of the latter deeper and further which will be shown in follows after studying this situation with a special method.

---

*Corresponding author: tgp6463@zjnu.cn

To discuss this behavior, we divide the Hamiltonian into three parts. The first one is given by the external field denoted by $H_e$, the second one is the spin-spin interaction we signed it as $H_s$, and the last one is the level-level interaction which only considers the levels in the neighborhood with each other and neglect others. The first one can be written as:

$$\begin{aligned}
H_e &= \frac{1}{8\pi v} \sum_k \sum_{\alpha=1,2} c^2 \frac{2\pi\hbar}{\omega} \varepsilon_k^{2(\alpha)} \cdot \\
&\int [\left| (\hat{a}_{k,\alpha}(t) \quad \hat{a}_{k,\alpha}^+(t)) \nabla \times \begin{pmatrix} u(k,r) \\ u^*(k,r) \end{pmatrix} \right|^2 \\
&+ \frac{1}{c^2} \left| (u(k,r) \quad u^*(k,r)) \frac{\partial}{\partial t} \begin{pmatrix} \hat{a}_{k,\alpha}(t) \\ \hat{a}_{k,\alpha}^+(t) \end{pmatrix} \right|^2 ] dV \\
&= \frac{1}{8\pi v} \sum_k \sum_{\alpha=1,2} c^2 \frac{2\pi\hbar}{\omega} \varepsilon_k^{2(\alpha)} \cdot \\
&\int [\left| 2[\text{Re}(a_{k,\alpha}(t)) \quad \text{Im}(\hat{a}_{k,\alpha}(t))] \nabla \times \begin{bmatrix} \text{Re}(u) \\ -\text{Im}(u) \end{bmatrix} \right|^2 \\
&+ \frac{1}{c^2} \left| 2[\text{Re}(u) \quad \text{Im}(u)] \frac{\partial}{\partial t} \begin{bmatrix} \text{Re}(\hat{a}_{k,\alpha}(t)) \\ -\text{Im}(\hat{a}_{k,\alpha}(t)) \end{bmatrix} \right|^2 ] dV
\end{aligned} \quad (1)$$

where $\varepsilon_k^{(\alpha)}$ are a polarization vector, and the direction of $(\varepsilon_k^{(1)}, \varepsilon_k^{(2)}, k/|k|)$ corresponds to the right-hand rule. The base vector $u(k,r)$ satisfies the normalizing condition

$$\frac{1}{V} \int u_{k,\alpha} \cdot u_{k',\alpha'} dV = \delta_{k,k'} \delta_{\alpha,\alpha'} \quad (2)$$

While the conjugate forms of $\hat{a}_{k,\alpha}^+$ and $u^*(k,r)$ make the multiply integration result $H$ as a real number. The operators $\hat{a}_{k,\alpha}^+$ and $\hat{a}_{k,\alpha}$ are given by

$$\begin{aligned}
\hat{a}_{k,\alpha}^+ &= (2\hbar\omega)^{1/2} (\omega \hat{Q}_{k,\alpha} - i\hat{P}_{k,\alpha}) \\
\hat{a}_{k,\alpha} &= (2\hbar\omega)^{1/2} (\omega \hat{Q}_{k,\alpha} + i\hat{P}_{k,\alpha})
\end{aligned} \quad (3)$$

and

$$\begin{aligned}
\hat{Q}_{k,\alpha} &= \frac{1}{c\sqrt{4\pi}} (C_{k,\alpha} + C_{k,\alpha}^*) \\
\hat{P}_{k,\alpha} &= \frac{-i\omega}{c\sqrt{4\pi}} (C_{k,\alpha} - C_{k,\alpha}^*)
\end{aligned}$$

where $C_{k,\alpha}$ and $C^*_{k,\alpha}$ are the coefficients of the Fourier Expansion for the vector potential in the initial time. So the $\hat{a}_{k,\alpha}(t)$ with its relation can be written as $[\hat{a}_{k,\alpha}(t), \hat{a}^+_{k,\alpha}(t)] = \delta_{kk'}\delta_{\alpha\alpha'}$, and the relationship between $\hat{P}_{k,\alpha}$ and $\hat{Q}_{k,\alpha}$ can be indicated by regular equation as follows

$$\frac{\partial H_e}{\partial Q_{k,\alpha}} = -\dot{P}_{k,\alpha}, \quad \frac{\partial H_e}{\partial P_{k,\alpha}} = \dot{Q}_{k,\alpha} \tag{4}$$

Above all, we can get the result of the Hamiltonian of a particle in *n* level by the action of external field can be written as a matrix form as follows

$$H_e(n) = I_n H_e \tag{5}$$

where $I_n$ is a unitary matrix, here we must point out that we suppose the external field which acting on the system is a homogeneous field. Thus the Hamiltonian $H_e(n)$ is only dependent with *n* level where the particle stays. Meanwhile, the spin-spin interaction Hamiltonian $H_s$ and level-level interaction Hamiltonian $H_i$ can be expressed, respectively

$$H_s = \lim_{\varepsilon \to 0} \int_{t-\varepsilon}^{t+\varepsilon} \omega(t)\sigma_{nz} d\varepsilon = \begin{Bmatrix} \omega(t)|\uparrow\rangle \\ \omega(t)|\downarrow\rangle \end{Bmatrix} \equiv E_n(t)_\pm \tag{6}$$

$$H_i = \lim_{h \to 0} \int_{t-h}^{t+h} [\Delta\varepsilon(n-1,n) - \Delta\varepsilon(n,n+1)] dt \equiv \overline{\Delta\varepsilon_n(t)} \tag{7}$$

where $\sigma_{nz}$ is Pauli matrix, and $\Delta\varepsilon(n, n+1)$ is the energy difference between the *n*th level and (*n*+1)th level, while the *n*th and (*n*+2)th are neglected. Note that the reduce form of $E_n(t)_\pm$ which stands for spin-spin interaction Hamiltonian is time dependent as well as the case of Hamiltonian $H_e$, however, the part $\overline{\Delta\varepsilon_n(t)}$ in Eq.(7) which shows the average interaction of particles in *n* level with its nearest neighbor levels is also time dependent, but we can recognize it as a form with time independent because its tiny changing with time, so it is equivalent of the form written as $\varepsilon_i$.

For the general case of an arbitrary linear Hamiltonian $H(t) = E + C(t)$, here *C* is a diagonal, there is no general solution yet, but for some survival probabilities, some exact results have been conjectured [7] and derived [8,9,23,24]. Especially in Ref.[7], Brundobler and Elser consider the formula for the certain diagonal elements of the *S* matrix for it obtains an approximation, that remains numerically very accurate when the approximation supposedly break down even at strong coupling. To make more accurate, when expand the diagonal matrix into the complex one which called three-diagonal one.

Based on the discussion shown in above, finally, it is naturally organized the Hamiltonian formula which represents all information of energy as a special matrix shown as follows

$$H=\begin{pmatrix} H_1 & \varepsilon_1 & 0 & 0 & \cdots & 0 \\ -\varepsilon_1 & H_2 & \varepsilon_2 & 0 & \cdots & 0 \\ 0 & -\varepsilon_2 & H_3 & \varepsilon_3 & \cdots & 0 \\ \vdots & \vdots & \ddots & \ddots & \ddots & \vdots \\ 0 & 0 & \cdots & -\varepsilon_{n-2} & H_{n-1} & \varepsilon_{n-1} \\ 0 & 0 & \cdots & 0 & -\varepsilon_{n-1} & H_n \end{pmatrix} \qquad (8)$$

where $H_{nn} = H_{en} + H_{sn} = H_{en} + \overline{E_n(t)_\pm}$ represents the sum of Hamiltonian which in the $n$th level both, while $\varepsilon_n = H_i = \overline{\Delta\varepsilon_n(t)}$ shown above indicates the interactions with its neighbor, unfortunately, it can not been added to the $H_n$ because it is seen as a constant discussed in above, unlike Hamiltonian $H_e$ and $H_s$

It is revealed by Eq.(8) that the Hamiltonian of a system can be expressed by a three-diagonal matrix, that is quite concise and convenient to deal with. As we know, all microscopic systems can be described by the $n$-dimensional Schrödinger equation

$$i\hbar\partial_t \psi_n(t) = H\psi_n(t) \qquad (9)$$

With the $\psi_n(t)$ given by $\psi_n(t) = [\psi_1(t) \quad \psi_2(t) \quad \cdots \quad \psi_n(t)]^T$.

V. N. Ostrovsky et al[25] worked for the multistate LZ theory assumes that the transitions between states can take place at the times at which the energies associated with two states cross, or have an avoided crossing according to the no crossing rule of von Neumann and Wigner. So the method they involved is to motivate such an assumption by the way of expanding the adiabatic basis states of the system shown as follows:

$$|\psi(t)\rangle = \sum_j c'_j \exp[-i\int_{t_i}^{t} \varepsilon_j(t')dt']|\chi_j\rangle \qquad (10)$$

It is a useful way to research for the LZ tunneling. But here, we connect another direction by dealing with the form of Hamiltonian $H(t)$ while the state function is just formed as above.

Therefore, we intuitively see that the main problem is how to transfer the Hamiltonian to solve the Eq.(9), thus the Hamiltonian shown in Eq.(8) can be transferred as

$$H=\begin{pmatrix} h_1 & 0 & 0 & \cdots & 0 \\ -\varepsilon_1 & h_2 & 0 & \cdots & 0 \\ 0 & -\varepsilon_2 & h_3 & \cdots & 0 \\ \vdots & \vdots & \ddots & \ddots & \vdots \\ 0 & 0 & 0 & -\varepsilon_{n-1} & h_n \end{pmatrix} \equiv \Gamma^{(h_n)}_{(-\varepsilon_{n-1})} \qquad (11)$$

where the intermediate parameters $h_n$ read as $h_1 = H_1$, $p_1 = \varepsilon_1^2/H_1$, $p_k = \varepsilon_k^2/h_k$, $h_{k+1} = H_k + p_k$ $(k = 2,3\ldots n)$. So we concisely rewrite the Hamiltonian $H$ as $H = \Gamma^{(h_n)}_{(-\varepsilon_{n-1})}$. After inserting it into Eq.(9), we can obtain that

$$\begin{cases} h_1\psi_1 = i\hbar\dot{\psi}_1 \\ -\varepsilon_{k-1}\psi_{k-1} + h_k\psi_k = i\hbar\dot{\psi}_k \end{cases} \quad (k=2,3\ldots n) \tag{12}$$

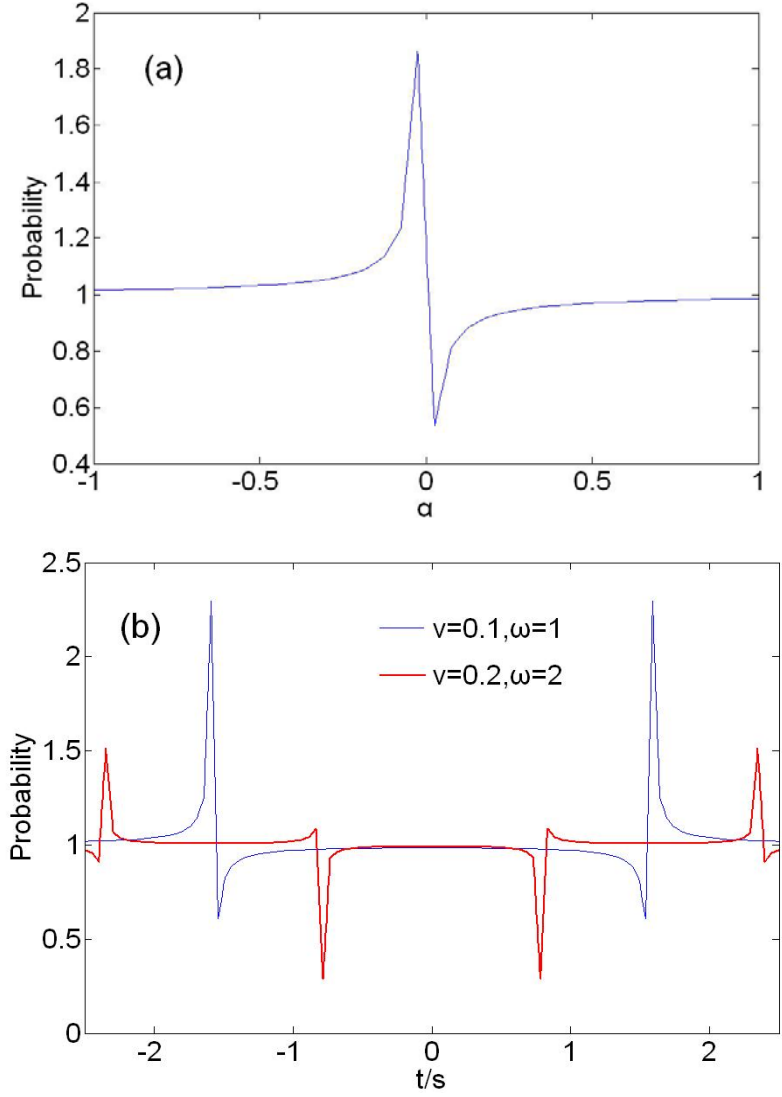

FIG.1 The Landau-Zener tunneling probability. (a) the classical case; (b) the transition probability varying with time in our case between two levels, where a comparison of $\omega=1$(blue line) and $\omega=2$(red line) with the same value of v(v=0.1) is given.

Note that based on the expression of $\psi$ by the solution of the upper formula in Eq.(12), we can insert it into the below formula with the case of $k=2$, and express the $\psi_2$, so, step by step, we can get the final expression of $\psi_i(t)$ with each $i$ case $(i=1,2,3\ldots n)$. At last, gets the result for the LZ tunneling probability P.

Without loss of generality, the Hamiltonian $H(t)$ in many cases can be transformed by Morris-Shore (MS) transformation [26], the form can be written as [27]

$$H(T) = SH(t)S^+ = \begin{bmatrix} 0 & V \\ V^+ & D(t) \end{bmatrix} \tag{13}$$

and

$$S = \begin{bmatrix} A & 0 \\ 0 & B \end{bmatrix}$$

$S$ is a constant transformation matrix which can be shown in the block matrix form, here $A$ and $B$ are the unitary square matrices with $N_a$-dimensional $N_b$-dimensional, respectively, and $V = AVB^+$. Here, the $H(t)$ matrix can be transformed an upper-triangular matrix because $A$ and $B$ are defined by the condition that they can diagonalize $VV^+$ and $V^+V$ respectively.[26]

The classical LZ model within the case of $\gamma(t) = \alpha t$ is the most basic process. According to the method discussed in this paper, we provide the external field $\gamma(t) = A\sin\omega t$ instead of $\gamma(t) = \alpha t$, and the transition probability in the adiabatic limit will shown as follows.

$$P = \exp(-\frac{\pi\varepsilon^2}{2A\omega\cos\omega t}) \tag{14}$$

Fig.1 shows the distinction of transition between the classical LZ model and our case. In classical LZ model, shown in Fig.1 (a), the probability will change with the value of $\alpha$, while $\alpha \to 0$, the transition probability between two levels break changes, out of the region, the P reaches steadily. But in our case, just as shown in Fig.1 (b), the probability changes periodically with the value of $t$, that is to say, in some special time, particle has high probability to transit to higher energy level, and then the system cuts down transition probability and reaches the adiabatic state with some particles locate in high level, in the next special time, other particles given the another chances and reach the higher state. As the time elapsed, the system gets up to the higher energy state with the help of the external field. By the way, the transition probability is also correlates with the value of $\omega$ (shown in Fig.1(b) with different colors). From the Fig.1(b), we can find that the probability amplitude will be smaller as well as the value of $\omega$ becomes bigger. That is to say, the probability amplitude will become larger in the case of the slow varying sweeping rate of external field.

With conclusion, we have studied the many body system in LZ tunneling, with $n$-level states although we consider the neighbor levels interaction only, the method of transferring the three-diagonal matrix to lower triangular one can be applied the five-diagonal matrix or even more as well, the difference between three-diagonal and five-diagonal matrix is that the former only consider the neighbor interaction but the later not just so, it takes the interaction with its two neighbor levels into consideration, of course, the case of the later is much complicated than the former, not only the calculation. Above all, this method could be applied theoretically in any case as long as its Hamiltonian can be transferred to lower triangular matrix.

In summary, we explore theoretically the effect of the varying sweeping rate of the external field upon the LZ tunneling by a precise method which has a wide application in physical process. The main significance of this paper has two: (a) find a new way to make the system transit to higher energy levels with the help of external field; (b) provide a concise method to solve the Schrödinger equation with the special Hamiltonian form which could describe many cases of physical models. For the future, we expect the concise method could be applied into physical experiment and make the study LZ tunneling further to interpret the more interesting quantum dynamical phenomena.

Fig.1

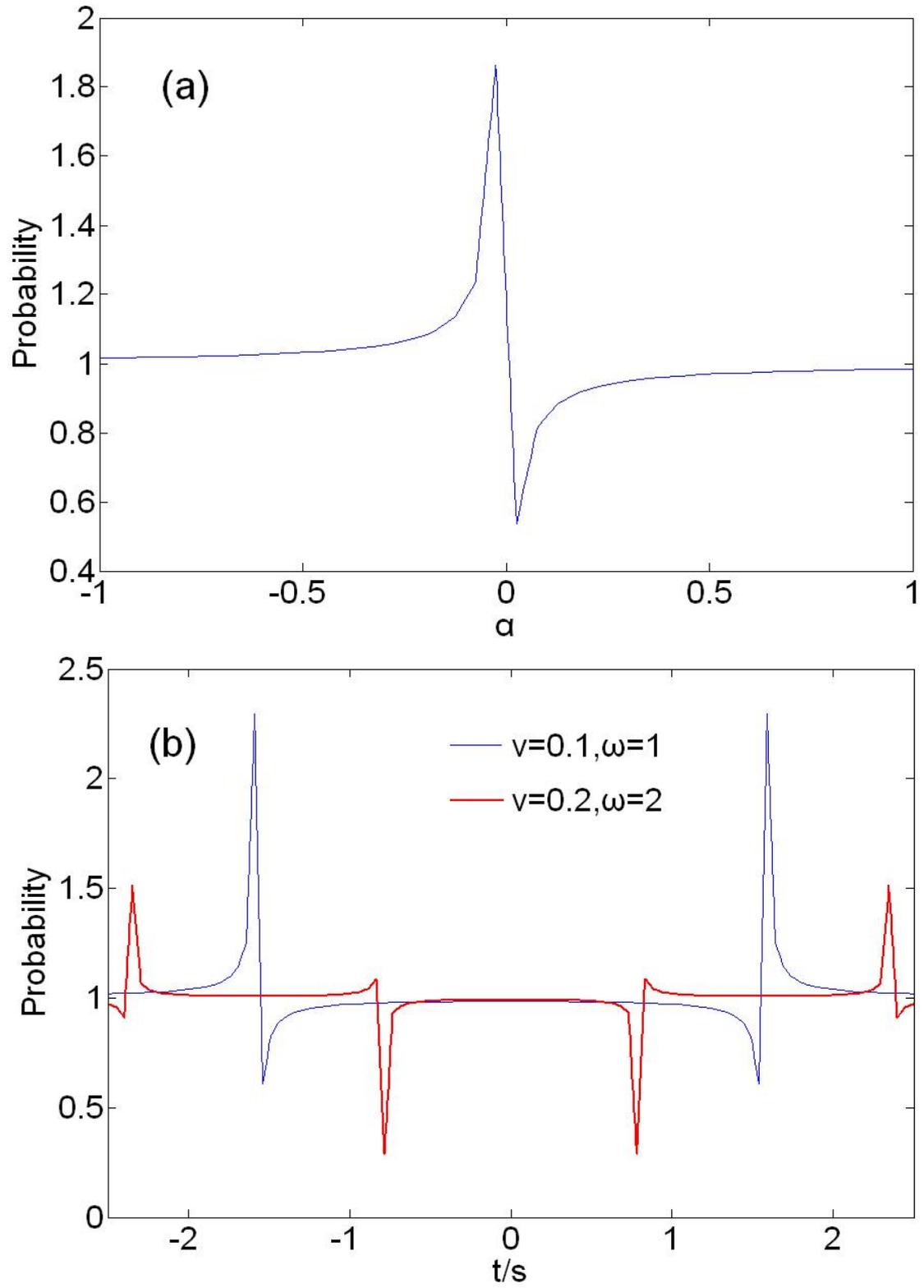